\documentclass{aa}

\usepackage{graphicx}
\usepackage{txfonts}
\usepackage{braket}
\usepackage{amsmath}
\usepackage{bm}
\usepackage[colorlinks,allcolors=blue]{hyperref}
\usepackage[a]{esvect}
\makeatletter
\renewcommand*\aa@pageof{, page \thepage{} of \pageref*{LastPage}} 

\makeatother
\usepackage{amstext}
\usepackage[normalem]{ulem}

\begin{document}

\title{Model Independent Periodogram for Scanning Astrometry}

\author{A. Binnenfeld\inst{\ref{inst1}} \and S. Shahaf\inst{\ref{inst2}} 
\and S. Zucker\inst{\ref{inst1}}}

\institute{Porter School of the Environment and Earth Sciences, Raymond and Beverly Sackler Faculty of Exact Sciences, 
Tel Aviv University, Tel Aviv, 6997801, Israel \\
\email{avrahambinn@gmail.com}\label{inst1}
\and
Department of Particle Physics and Astrophysics, Weizmann Institute of Science, Rehovot 7610001, Israel
\label{inst2}}

\date{Accepted XXX. Received YYY}

\abstract{
We present a new periodogram for periodicity detection in one-dimensional time-series data from scanning astrometry space missions, like Hipparcos or Gaia. The periodogram is non-parametric and does not rely on a full or approximate orbital solution. Since no specific properties of the periodic signal are assumed, the method is expected to be suitable for the detection of various types of periodic phenomena, from highly eccentric orbits to periodic variability-induced movers. The periodogram is an extension of the phase-distance correlation periodogram (PDC) we introduced in previous papers based on the statistical concept of distance correlation. We demonstrate the performance of the periodogram using publicly available Hipparcos data, as well as simulated data. We also discuss its applicability for Gaia epoch astrometry, to be published in the future data release 4 (DR4). 
}


\titlerunning{Periodogram for Scanning Astrometry}
\authorrunning{A. Binnenfeld et al.}

\maketitle
\section{Introduction}
\label{sec:intro}


Major astrometric survey space missions such as Hipparcos and Gaia provide position and parallax measurements for a large number of celestial objects \citep{hipcat97, gaia2016}. 
Both missions monitor the sky following a complex scanning law, yielding a sparse unevenly-sampled astrometric time series of the various targets \citep{vanev98, gaia2016}. This scanning is performed along reference great circles (RGC), gradually covering the entire celestial sphere. Thus, in each epoch, an object location is conveyed by two quantities: the \textit{scanning angle} $\theta$, and the \textit{abscissa} $v$. 

\begin{figure} 
\includegraphics[width=0.95\linewidth,clip=true]{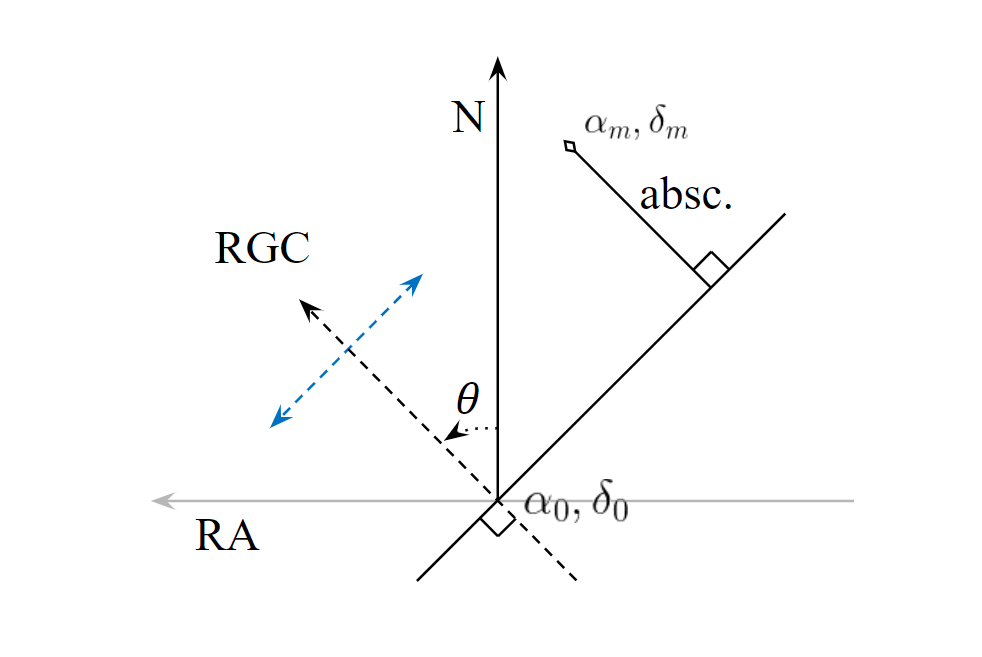}
\centering
\caption{Definition of the abscissa residual $v$ for a single-epoch position ($\alpha_m$, $\delta_m$), with respect to the nominal catalogue position ($\alpha_0$, $\delta_0$) and the RGC orientation $\theta$. The actual position of the observed target lies somewhere along the blue dotted arrowed segment. The figure is based on one originally published by \citet{vanev98}.}
\label{fig:ANG}
\end{figure}

In its original meaning, the Latin word `abscissa' refers to the horizontal coordinate in a two-dimensional Cartesian system. In the context of scanning astrometry, the word abscissa refers to one-dimensional positional coordinates that are measured along the scanning direction of the instrument. While scanning astrometry provides  well-constrained abscissae, positional measurements in the cross-scan direction, perpendicular to the scanning direction, are obtained with significantly lower precision. As a result, scanning astrometry data are often considered one-dimensional  positional measurements. 
The information they convey is only partial information regarding the astrometric position, while the actual position lies somewhere on an imaginary line perpendicular to the scanning direction. Note that the abscissa is often specified with respect to the catalogue nominal position. Fig.~\ref{fig:ANG} illustrates the geometric setting described above.

Full astrometric solutions for the observed targets include several parameters fitted to minimize the model errors in terms of the abscissa residuals \citep[see][]{hipcat97, ESAcat97}.  In the simplest cases, the basic five-parameter solution (right ascension, declination, two-dimensional proper motion vector, and parallax) is sufficient to model the target's proper motion combined with the apparent motion induced by the motion of Earth. When considering the astrometric motion of a binary system, seven or nine parameters are required to account for the target's acceleration. In some cases, if the orbital period is sufficiently short, a twelve-parameter model is fitted, to fully account for the Keplerian motion of the system. Solving for those parameters is more complicated and might even be quite computationally demanding and prone to errors of various kinds \citep[e.g.][]{vl2007new}. 

With the very high angular precision of modern instruments such as Gaia, the situation might become even more complicated when the binary system includes compact objects, and relativity comes into effect through light deflection and delays \citep[e.g.][]{Halbwachs2009}. Apparent periodic astrometric motion can also occur in unresolved visual double stars, in which one of the components is a periodically pulsating star. In these cases, commonly known as VIMs (Variability Induced Movers), the photocenter of the two blended stars undergoes a periodic displacement, which is related to the photometric periodic variability \citep{VIMS96}.

In other types of astronomical observations, such as radial velocities (RV) or photometry, the measured quantity is simply a scalar. The common procedure in this context is to first look for periodicity. If such periodicity is indeed found and a crude estimate of the period is obtained, then a full characterization of the variability is attempted, such as a Keplerian RV model, an eclipsing-binary model, a Cepheid light-curve model, etc. The most common approach to the first step, detecting the periodicity, is using a \textit{periodogram}. A periodogram scans a grid of trial periods (or frequencies) and assigns each one a score. This score quantifies the plausibility that a periodicity exists in the data, assuming this particular trial period. 

The most commonly used periodogram for scalar data is probably the \textit{Generalized Lomb-Scargle} (GLS) periodogram, which fits a sinusoidal function to the data and quantifies the goodness of fit \citep{GLSpaper}. The underlying assumption is that any periodic function can be described as a sum of harmonics, and assuming the first harmonic is the dominant one, the function can be approximated by a sinusoid. 
However, the case of abscissa data resulting from scanning astrometry is more complicated and requires special treatment. This is mainly related to the fact that every abscissa is measured along a different scanning direction. Therefore, the data cannot be treated in the same way as simple scalar data.

In this work, we present a non-parametric method to detect periodicity in time-series of one-dimensional astrometric data. This periodogram does not search for a specific type of periodic astrometric motion, and it is based on the concept of \textit{phase distance correlation} (PDC). \citet{zucker18} introduced the PDC periodogram as a new method to detect periodicity in time-series data. Essentially, for each trial period, PDC quantifies the statistical dependence between the measured quantity and the phase (according to the trial period), using the recently introduced distance correlation \citep{Szeetal2007}.

Distance correlation is a measure of the statistical dependence between random variables (as opposed to the correlation coefficient, which is a measure of linear dependence). One of its merits is that the two random variables need not be of the same dimensions and can even be drawn from general metric spaces, not necessarily Euclidean \citep{Lyons2013}. \citeauthor{Lyons2013} showed that if the two metrics involved are of \lq strong negative type\rq \citep*[see][]{Zinger1992ACO}, distance correlation is guaranteed to be applicable as a measure of statistical dependence. 

The PDC periodogram, as introduced by \citet{zucker18}, calculates for each trial period, the distance correlation between the analyzed quantity (such as RV) and the phase according to the trial period. 
As Euclidean spaces are themselves of strong negative type \citep{Lyons2013}, \citet{zucker19} further formulated and demonstrated the application of the PDC periodogram to two-dimensional astrometry, using the regular Euclidean metric to quantify distances between different measurements.

In the next section, we discuss the specific adaptation of the PDC concept to scanning astrometry. Sect.~\ref{sec:ex2} demonstrates the application of the new periodogram to Hipparcos data, as well as additional simulated data. We conclude in Sect.~\ref{sec:conc} and discuss the method and its potential for future studies, particularly in the context of Gaia data.

\section{PDC for one-dimensional scanning astrometry}
\label{sec:mtric}

\subsection{A new metric}

In order to extend the PDC concept to data from scanning astrometry, we had to define a suitable metric for the set of observations, using the above-mentioned parameters: scanning angle $\theta$ and abscissa $v$. This metric should then enable estimating the distance correlation measure used to compute the PDC periodogram. After experimenting with a few alternative metrics, we chose the following, which is based on the concept of \textit{Energy Distance}.

Energy distance is a statistical distance between probability distributions, first introduced by G\'{a}bor Sz\'{e}kely \citep{Szkely2013EnergySA}. Note that energy distance is a metric of strong negative type \citep{Rizzo2016}, and therefore satisfies the sufficient condition for the applicability of distance correlation as an independence test, as mentioned in the previous section.

As shown in Sect.~\ref{sec:intro}, the actual astrometric position of the target can be assumed to lie on a line segment whose length is constrained by the width of the scanning sensor (Fig.~\ref{fig:ANG}). We do not have a reason to assume that there is a preference for one particular area of the detector over another, and therefore all points on the segment are a priori equally probable. 

In a Bayesian inference setting, we could set a uniform prior for the position along the width of the detector. However, our approach is based on distance correlation and therefore relies on our ability to define a metric between pairs of measurements. To qualitatively reflect the logic of an uninformative prior, we choose to represent each segment only by the position of its edges. This choice defines a one-to-one correspondence between the set of all segments and the set of all unordered pairs of points on the plane.

For the set of all pairs of points on the plane we can use the Energy Distance metric, in the following way: let  $U$ and $W$ be two segments in the plane, which can be represented by their endpoints: $\boldsymbol{u_1}$, $\boldsymbol{u_2}$ and $\boldsymbol{w_1}$,$\boldsymbol{w_2}$, respectively. We can then consider each segment as a sample of two points and calculate the energy distance between the two samples corresponding to $U$ and $W$ \citep{Szkely2013EnergySA}:
\begin{equation}
\mathcal{D}(U,W) = \frac{1}{2} \sum_{i=1}^2 \sum_{j=1}^2 {|\boldsymbol{u_i} - \boldsymbol{w_j}|} - \frac{1}{2} |\boldsymbol{u_1} - \boldsymbol{u_2}| - \frac{1}{2} |\boldsymbol{w_1} - \boldsymbol{w_2}|
\label{en_eq}
\end{equation}
Note that $|\boldsymbol{u_1} - \boldsymbol{u_2}|$ is actually the length of the segment $U$. Now, let the two segments be defined by the abscissae $v_1$ and $v_2$ and the scanning directions $\theta_1$ and $\theta_2$ (see Fig.~\ref{fig:ANG}), and assume that the width of the scanning field of view, and therefore the length of the segments, is $L$. The segments are then uniquely defined by the following two sets of vectors on the plane:
\begin{equation}
\begin{gathered}
\label{eq:vectors}
U = \left\{
\begin{pmatrix} v_1 \sin{\theta_1} - \frac{1}{2} L \cos{\theta_1}\\ v_1 \cos{\theta_1} + \frac{1}{2} L \sin{\theta_1}\\ \end{pmatrix},
\begin{pmatrix} v_1 \sin{\theta_1} + \frac{1}{2} L \cos{\theta_1}\\ v_1 \cos{\theta_1} - \frac{1}{2} L \sin{\theta_1}\\ \end{pmatrix}
\right\} \\ \\ 
W = \left\{
\begin{pmatrix} v_2 \sin{\theta_2} - \frac{1}{2} L \cos{\theta_2}\\ v_2 \cos{\theta_2} + \frac{1}{2} L \sin{\theta_2}\\ \end{pmatrix},
\begin{pmatrix} v_2 \sin{\theta_2} + \frac{1}{2} L \cos{\theta_2}\\ v_2 \cos{\theta_2} - \frac{1}{2} L \sin{\theta_2}\\ \end{pmatrix}
\right\}
\end{gathered}
\end{equation}
We can now substitute the endpoints of the segments defined in Eq.~\ref{eq:vectors} into the definition of the energy distance (Eq.~\ref{en_eq}), and get the following expression:
\begin{multline}
\mathcal{D}(U, W) =
\frac{1}{2} \biggl(\, \Bigl\{ v_1^2 - 2v_1v_2 \cos(\theta_1 - \theta_2) + v_2^2 \\ 
+ L (v_1 + v_2)\sin(\theta_1 - \theta_2)
+ \frac{1}{2} L^2 \Bigl[\,1 + \cos(\theta_1 - \theta_2)\Bigr]\,\Bigr\}^\frac{1}{2} \\
+ \Bigl\{(v_1^2 - 2v_1v_2 \cos(\theta_1 - \theta_2) + v_2^2 \\
+ L (v_1 - v_2)\sin(\theta_1 - \theta_2)
+ \frac{1}{2} L^2 \Bigl[\,1 - \cos(\theta_1 - \theta_2)\Bigr]\,\Bigr\}^\frac{1}{2} \\
+ \Bigl\{v_1^2 - 2v_1v_2 \cos(\theta_1 - \theta_2) + v_2^2 \\
- L (v_1 - v_2)\sin(\theta_1 - \theta_2)
+ \frac{1}{2} L^2 \Bigl[\,1 - \cos(\theta_1 - \theta_2)\Bigr]\,\Bigr\}^\frac{1}{2} \\
+ \Bigl\{(v_1^2 - 2v_1v_2 \cos(\theta_1 - \theta_2) + v_2^2 \\
- L (v_1 + v_2)\sin{(\theta_1 - \theta_2)}
+ \frac{1}{2} L^2 \Bigl[\,1 + \cos(\theta_1 - \theta_2)\Bigr]\,\Bigr\}^\frac{1}{2} \biggr)\, 
-L
\label{dist_eq}
\end{multline}

Thus we have obtained a metric for the set of astrometric measurements, without any further assumptions concerning the observed one-dimensional object position (i.e. model-independent). We will use it in the following section to formulate the PDC for one-dimensional astrometric measurements. 

In developing the above metric we assumed there is no information available about the location of the object in the direction perpendicular to the scanning direction. While some negligible information might be effective, we preferred to follow the guidelines proposed by \citet{LindegrenBastian} who recommended treating measurements of space scanning astrometry as one-dimensional.

\subsection{Astrometric PDC periodogram formulation}

Following \citet{zucker18} and \citet{zucker19}, let us define a distance matrix based on the metric we have introduced in Eq.~\ref{dist_eq}. For each pair of astrometric measurements ($i$ and $j$), the entry in the distance matrix would then be:
\begin{equation}
a_{ij} = \mathcal{D}(i, j) \ .
\end{equation}

For each trial period $P$ we define a phase distance matrix:
\begin{equation}
\begin{split}
\phi_{ij} &= (t_i - t_j)\mod P \ , \\
b_{ij} &= \phi_{ij}(P-\phi_{ij}) \ .
\end{split}
\end{equation}

Now we apply $\mathcal{U}$-centering to the two matrices, which will allow an unbiased estimator of the distance correlation \citep{sz_par_14}:
\begin{equation}
A_{ij} = \begin{cases}
\begin{split}
a_{ij} - \frac{1}{N-2}\sum\limits_{k=1}^{N}a_{ik} 
- \frac{1}{N-2}\sum\limits_{k=1}^{N}a_{kj} \\
+ \frac{1}{(N-1)(N-2)}\sum\limits_{k,l=1}^{N}a_{kl} 
\end{split} 
& \text{if $i \neq j$ ,} \\ \\
0 & \text{if $i = j$ .}
\end{cases}
\end{equation}

A similar procedure is applied to obtain the matrix $B_{ij}$ from $b_{ij}$. Using the $\mathcal{U}$-centered matrices the unbiased estimator of the distance correlation can now be computed via the expression:
\begin{equation}
\label{eq:cor}
D = \frac{\sum\limits_{ij}A_{ij}B_{ij}}{\sqrt{(\sum\limits_{ij}A^2_{ij})(\sum\limits_{ij}B^2_{ij})}} \ .
\end{equation}

The significance of a period detected by PDC can be assessed using a permutation test, which we have indeed used in previous papers  \citep{binnenfeld20, Binnenfeld2022}: we created a sample of $D$ values by randomly shuffling the assignment of measurements to phases, and recalculating $D$ for this random allocation of phases. Then we could obtain a threshold value matching a desired level of false-alarm probability (FAP).

In this paper, we use instead a chi-square test tailored for distance correlation, which was recently proposed by \citet{fap_shen}. It is very fast to compute and it spares the need for the computationally heavy permutation test.

\section{Examples}
\label{sec:ex2}

\subsection{Hipparcos intermediate astrometric data}

The launch of the Hipparcos Space Telescope in 1989 was a major breakthrough in astronomy. In its four-year mission it measured positions, parallaxes, and proper motions for about 120,000 stars, and its precision of up to a few mili-arc-seconds (mas) was unprecedented at the time. The Hipparcos Catalogue was published in 1997 \citep{hipcat97}.

To test and demonstrate our newly developed periodogram, we used the publicly available Hipparcos intermediate astrometric data (IAD), which include the individual abscissa residuals (after subtracting the basic astrometric model) for each source in the catalogue, and are publicly available. Two different versions of the data are offered, produced by the two different data analysis consortia FAST \citep{FAST} and NDAC \citep{NDAC}.

We have experimented with various possible values of $L$, the length of the segments on which the actual astrometric position of the target can be assumed to lie. We found that values smaller than a few milli-arc-seconds produced very similar periodograms. Thus, we opted to use $L = 1$ mas in our demonstrations.

\subsubsection{Binary stars}

\begin{figure*} 
\centering
\includegraphics[width=1.6\columnwidth,clip=true]{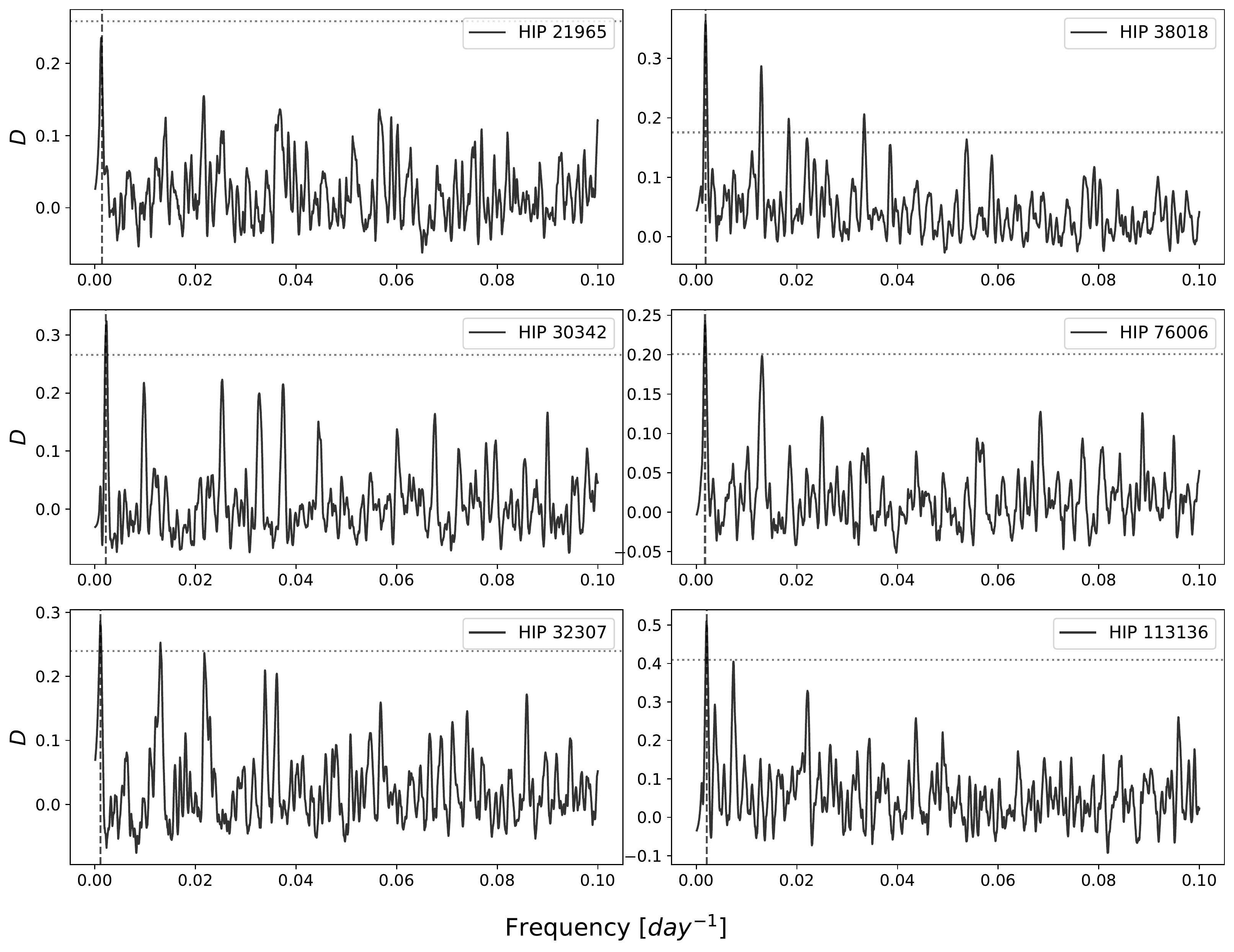}
\caption{Astrometric PDC periodograms for selected Hipparcos binary-star targets from the catalogue of astrometric orbits published by \citet{Goldin_2007}. More information on the targets can be found in Table~\ref{tbl:bsS_param}. The dashed vertical lines mark the published  orbital frequencies, and the dotted horizontal lines correspond to a FAP level of $10^{-3}$.}
\label{fig:bsS}
\end{figure*}

To demonstrate the capability of our astrometric PDC method to detect periodic orbits of binary stars, we used a catalogue of astrometric orbits from Hipparcos published by \citet{Goldin_2007}. The stars included in this catalogue had been classified in the original Hipparcos catalogue as 'stochastic binaries', i.e., binary stars whose orbits could not have been characterized by the Hipparcos teams.

Fig. \ref{fig:bsS} presents six example periodograms selected out of many positive results for the catalogue items, all based on NDAC data. In all of the PDC periodograms we obtained, a significant peak appears at the expected frequency, corresponding to the period of the published solution (marked in dashed vertical lines). We also mark in dotted horizontal lines the value corresponding to a FAP level of $10^{-3}$. More information on the presented targets can be found in Table~\ref{tbl:bsS_param}.

   \begin{table*}
      \caption[]{Details of the binary stars used for demonstration (Fig.~\ref{fig:bsS}).}
     $$ 
         \begin{array}{llcccc}
            \hline
            \noalign{\smallskip}
            \textrm{HIP} & \textrm{Designations} & \textrm{\# epochs} &\textrm{Published period} & \textrm{Eccentricity} & \textrm{PDC period} \\
             & & & \textrm{[days]} &  &\textrm{[days]}\\
            \noalign{\smallskip}
            \hline
            \noalign{\smallskip}
            \textrm{HIP 21965} & \textrm{HD 30051} & 38 & 709^{+26}_{-23}   & 0.29^{+0.30}_{-0.17}  & 746 \\
            \noalign{\smallskip}
            \textrm{HIP 30342} & \nu\,\textrm{ Pictoris} & 37&452^{+13}_{-16}   & 0.20^{+0.35}_{-0.19}   & 435 \\
            \noalign{\smallskip}
            \textrm{HIP 32307} & \textrm{HD 49377} & 41 & 889^{+53}_{-56}  & 0.15^{+0.25}_{-0.16}   & 909 \\
            \noalign{\smallskip}
            \textrm{HIP 38018} & \textrm{HD 61994} & 56 & 547^{+8}_{-9}  & 0.52^{+0.11}_{-0.09}   & 549 \\
            \noalign{\smallskip}
            \textrm{HIP 76006} & \textrm{HD 138525} & 49
 & 572^{+21}_{-21}  & 0.51^{+0.32}_{-0.27}  & 588  \\
            \noalign{\smallskip}
            \textrm{HIP 113136} & \delta \textrm{ Aquarii} & 24 & 483^{+20}_{-19}  & 0.12^{+0.25}_{-0.15} & 485 \\
            \noalign{\smallskip}
            \hline
         \end{array}
     $$ 
    \label{tbl:bsS_param}
   \end{table*}

\subsubsection{Brown dwarfs}

\begin{figure*} 
\centering
\includegraphics[width=1.6\columnwidth,clip=true]{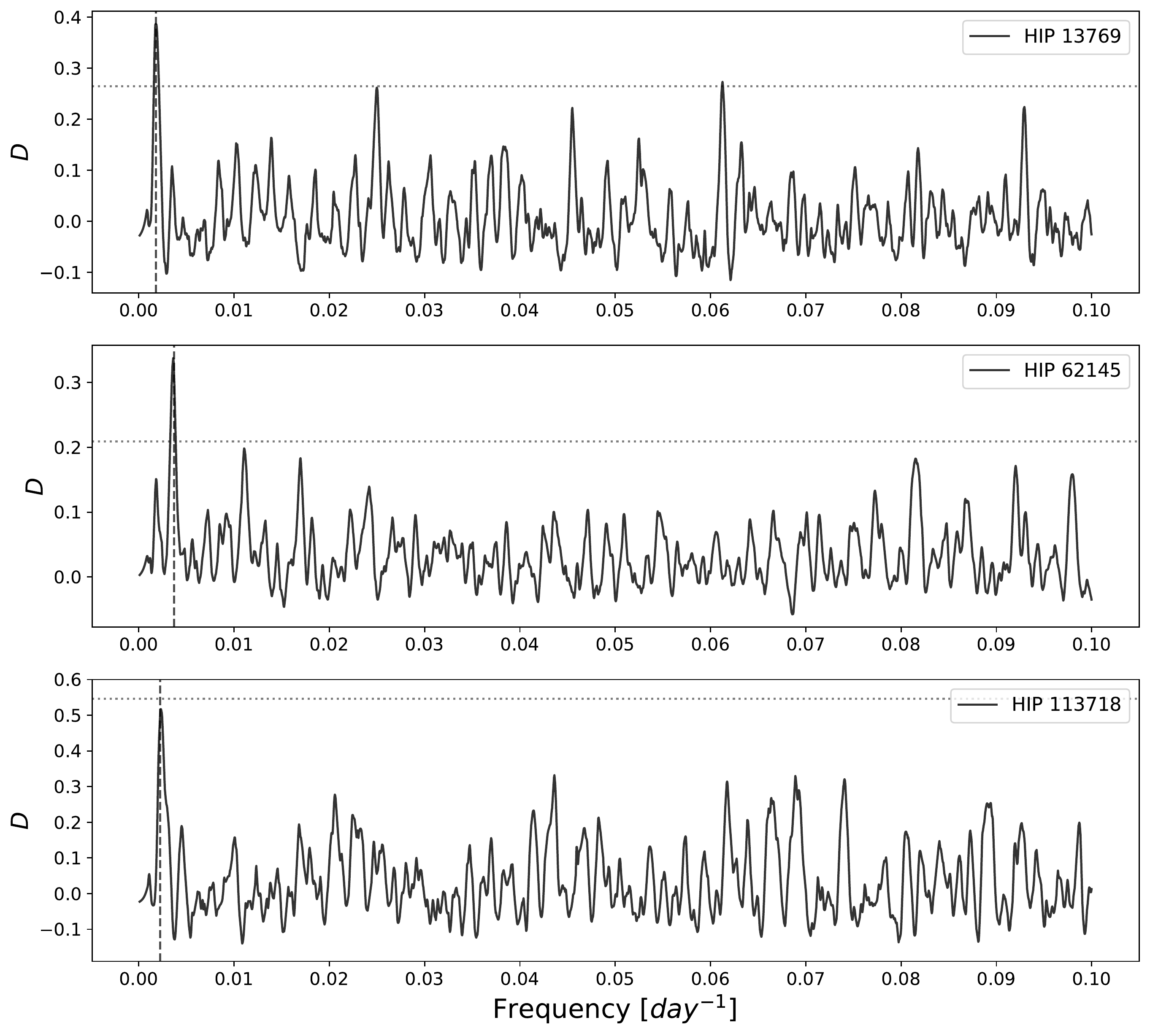}
\caption{Astrometric PDC periodograms for selected brown-dwarf hosting stars. More information on the targets can be found in Table~\ref{tbl:bdS_param}. The dashed vertical lines mark the published orbital frequencies, and the dotted horizontal line corresponds to a FAP level of $10^{-3}$.}
\label{fig:bdS}
\end{figure*}

Following the positive results obtained for binary-star systems, we found our method to be sensitive enough to detect astrometric orbits induced by brown-dwarf companions as well.

Fig.~\ref{fig:bdS} presents three examples of the results obtained by applying our method to the Hipparcos IAD of known brown-dwarf hosting stars. We applied it to the FAST data of HIP\,13769 and HIP\,113718, and to the NDAC data of HIP\,62145. Their orbital parameters, listed in Table~\ref{tbl:bdS_param}, were published by \citet{Tokovinin94}, \citet{Halbwachs2000}, and \citet{Reffert11}.

In all of the presented periodograms pertaining to those candidates, a significant peak appears close to the expected frequency, matching the orbital period of the known companion (marked in dashed vertical lines). FAP levels of $10^{-3}$ are marked in dotted horizontal lines.

   \begin{table*}
      \caption[]{Details for the brown-dwarf hosting stars used for demonstration.}
     $$ 
         \begin{array}{llccccc}
            \hline
            \noalign{\smallskip}
            \textrm{HIP} & \textrm{Designations} & \textrm{\# of epochs} & \textrm{Published period} & \textrm{eccentricity} & \textrm{Min. mass} & \textrm{PDC period}  \\
             &  & & \textrm{[days]} &  & [M_\textrm{J}] & \textrm{[days]} \\
            \noalign{\smallskip}
            \hline
            \noalign{\smallskip}
            \textrm{HIP 13769} & \textrm{HD 18445} & 26 & 554.58^{+1.25}_{-1.25}   & 0.558^{+0.067}_{-0.067}  & 44  & 562 \\
            \noalign{\smallskip}
            \textrm{HIP 62145} & \textrm{HD 110833} & 
 47 & 572^{+21}_{-21} & 0.51^{+0.32}_{-0.27} & 17 & 588  \\
            \noalign{\smallskip}
            \textrm{HIP 113718} & \textrm{HD 217580} & 18 &454.66^{+0.94}_{-0.94}  & 0.52^{+0.022}_{-0.022} & 67 & 442.5 \\
            \noalign{\smallskip}
            \hline
         \end{array}
     $$ 
    \label{tbl:bdS_param}
   \end{table*}

\subsection{Simulated data}

\begin{figure*} 
\centering
\includegraphics[width=1.6\columnwidth,clip=true]{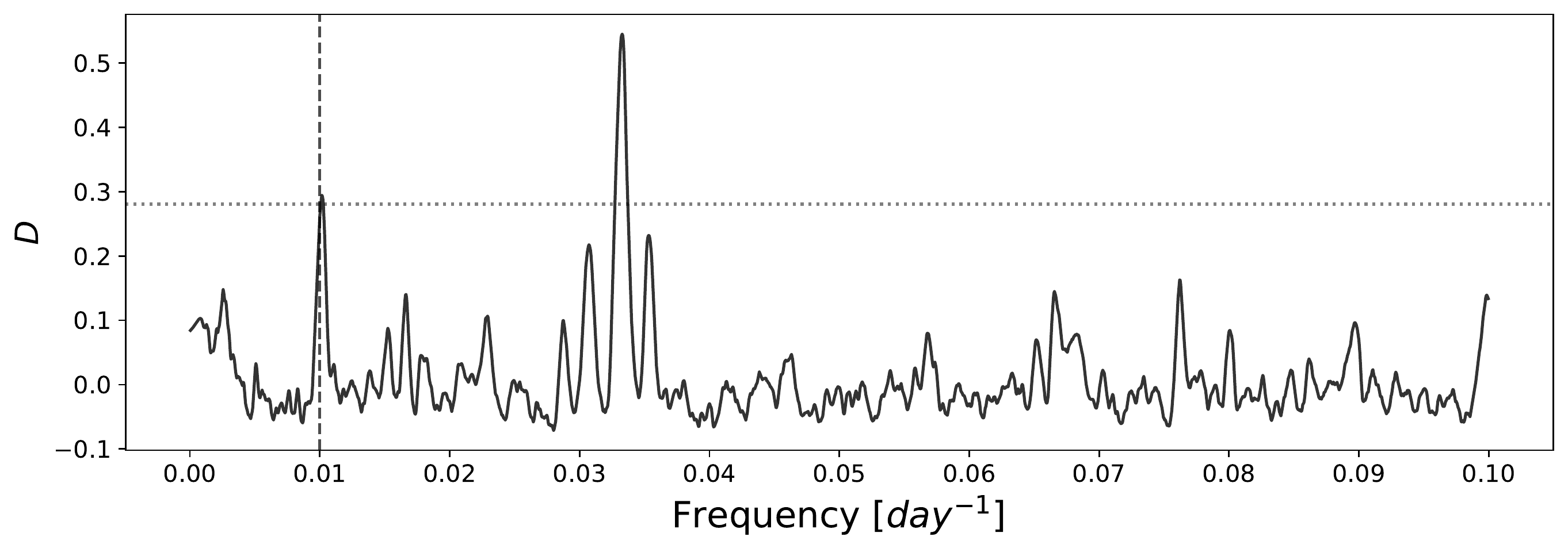}
\caption{Astrometric PDC periodograms for a simulated system with a 100-day orbital period and a 60-day satellite rotation period. The dashed vertical line marks the simulated orbital frequency, and the dotted horizontal line corresponds to a FAP level of $10^{-3}$. Note that an additional peak matching the period of $\sim30$ days appears in the periodogram, and it is probably related to half the satellite rotation period.}
\label{fig:sim_1}
\end{figure*}

The above results demonstrate that our newly developed periodogram is effective for periodicity detection in Hipparcos IAD. The data are given with respect to the catalogue position, and therefore they are corrected for the proper motion and parallactic motion of the observed star. However, in principle, especially in the presence of some astrophysical periodicity, the correction might not be perfect, and some residual effects of the parallax and the proper motion might still be present. To demonstrate the periodogram suitability in those conditions, we have further tested it using a simulation. In addition, the scanning law of the telescope might be affected by the motion of the satellite and as a result, the scanning angle might not be completely random. We included such an effect in our simulation as well.

We randomly drew $35$ epochs from a uniform distribution over an interval of three years. We used them to sample the astrometric signature of a Keplerian orbit with a semi-major axis of $5\,\textrm{mas}$, high eccentricity ($e = 0.8$), and a period of $100$ days. We added to it a circular $5\,\textrm{mas}$ orbit with a period of $365$ days and a $60^{\circ}$ inclination, to simulate a typical parallax signal, and also a linear trend simulating the effect of a linear $3\,\textrm{mas\,year}^{-1}$ proper motion. We further added white noise to the data at a signal-to-noise ratio of $\textrm{SNR} = 75$.
The astrometric position at each epoch was then translated into the scanning angle and abscissa representation, with the scanning angle following a simulated $60$ days satellite rotation period. 

Fig.~\ref{fig:sim_1} shows the results of applying the astrometric PDC to the simulated data described above. A prominent peak indeed appears at the simulated orbital period. An additional peak matching a $\sim30$\,day period appears in the periodogram, and it is probably due to interaction between the proper motion and the sampling window function caused by the satellite simulated 60-day rotation period. The residual parallax and proper motion, at the levels we introduced them, do not seem to affect the periodogram ability to detect the orbital period of the system, though a significant peak does appear, as might be expected, around a one-year period, probably related to the residual parallactic motion.

\subsection{Comparison with \citet{delisle2022analytical}}

In a recent publication, \citet{delisle2022analytical} presented an astrometric periodogram based on a linearized Keplerlian model fitted to a set of one-dimensional astrometric measurements. Their approach relies on analytical approximations and, therefore,  is probably superior to ours in terms of its computational cost. They also show that the statistical efficiency of their method degrades slowly as the orbital eccentricity increases, so that a ${\sim}10\%$ loss in power is only achieved for $e\,{\gtrsim}\,0.8$. Nevertheless, their approach is still model-dependent; for example, it might be suboptimal for detecting extremely eccentric orbits or variability-induced motion. When considering the wide range of variability sources, astrophysical or otherwise, the non-parametric approach presented in this work might prove valuable, and complementary to other techniques.

\begin{table*}
  \caption[]{Comparison of the period detection by \cite{delisle2022analytical} to those performed using the astrometric PDC periodogram for HIP 117622 and HIP 12726. We used the literature parameters published in \citet{Goldin_2007}.}
     \label{table:comp}
 $$ 
     \begin{array}{lccccccccc}
        \noalign{\smallskip}
         \multicolumn{1}{l}{\text{HIP}} & 
         \multicolumn{1}{l}{a_0\text{ [mas]}} & 
         \multicolumn{1}{c}{e} & 
         \multicolumn{1}{l}{\textrm{literature period [d]}} & 
         \multicolumn{1}{l}{\textrm{D\&S period [d]}} & 
         \multicolumn{1}{l}{\text{D\&S FAP}} & 
         \multicolumn{1}{c}{\textrm{PDC period [d]}}  &
         \multicolumn{1}{c}{\text{PDC FAP}} \\
        \noalign{\smallskip}
        \hline
        \noalign{\smallskip}
        \textrm{117622} &  12.1^{+2.2}_{-1.2} &  0.25^{+0.25}_{-0.16} & 1008^{+161}_{-41} & 1049 & 3 \times 10^{-5} & 943 & 5 \times 10^{-3} \\ \\
        \textrm{12726} & 11.7^{+3.8}_{-1.6} &  0.61^{+0.16}_{-0.13} & 536^{+12}_{-12} & 570 & 1 \times 10^{-3} & 572 & 1 \times 10^{-4} \\
        \noalign{\smallskip}
        \hline
        \noalign{\smallskip}
     \end{array}
 $$ 
\end{table*}

To demonstrate their method, \citet{delisle2022analytical} used two systems from the Hipparcos database that exhibited astrometric periodicity: HIP 117622 and HIP 12726. We compared the results they obtained in detecting the periodic signals known from the literature to those of the astrometric PDC periodogram. As can be seen in Table \ref{table:comp}, the astrometric PDC periodogram successfully detects the known periodicities, with corresponding FAP values comparable to those of \citet{delisle2022analytical} for the moderately eccentric HIP 117622 ($e = 0.25$), and an order of magnitude smaller for the more eccentric HIP 12726 ($e = 0.61$).

\section{Conclusion}
\label{sec:conc}

In this paper, we have presented the astrometric PDC method and demonstrated that it can be used to detect and quantify periodicity in the Hipparcos catalogue. It offers a new approach to studying data from scanning astrometry while avoiding complex, model-dependent solutions characterizing the variability, such as Keplerian models. We believe this approach can pave the way to new discoveries and insights.

Considered by many to be Hipparcos successor, Gaia was launched in 2014, with its latest data release including more than $1.8$ billion stars with an unparalleled high precision \citep[DR3,][]{gaia22}.

The epoch astrometry data of Gaia is expected to become available to the community as part of the future data release 4 (DR4). Previous data releases have already shown its potential in the classification of stellar multiplicity \citep{gaia_mult, gaia_mult_ast}, and also in the detection of substellar and planetary mass companions \citep{dr3holl}. Therefore, we expect our periodogram to be useful for the analysis of the Gaia astrometric data, potentially discovering signals that cannot be identified using standard analysis methods.

In a recent paper \citep{Binnenfeld2022}, we have introduced the \textit{partial phase distance correlation periodograms}, which allow accounting for 'nuisance' parameters in order to eliminate spurious peaks related to it. In the case of one-dimensional astrometry, the angle $\theta$ describing the RGC orientation may be considered a nuisance, since it is determined arbitrarily by the satellite orbit, and not related to the examined astrophysical phenomenon \citep[see][]{HOLL22}. Therefore, we plan to explore reducing its effect using the partial distance correlation periodograms. 

Using Hipparcos photometry as a nuisance variable to the astrometry may be informative as well. It might help in distinguishing orbital movement from luminosity variability in the case of VIMs. Gaia DR4 will also include various other types of epoch data, such as radial velocities and photometry in different bands, which we may be able to use in a similar way.

Hipparcos data and its astrometric solutions were extensively researched in search of stellar multiplicity, and also for orbital companions such as brown dwarfs and massive planets \citep[e.g.][]{hipold99, hipold2000, hipold2001, sode2010, snebrow18}. Nevertheless, we believe that due to the unique qualities of our new method, using it to carefully analyze the Hipparcos catalogue in its entirety can potentially reveal many new astrometric orbits, especially those of high eccentricity.
 
Another potential context in which our newly developed tool can contribute significantly is the Nancy Grace Roman Space Telescope \citep[NGRST, known before as WFIRST;][]{NGRST13}, currently expected to launch by 2027. 

We provide our Python implementation of the periodogram in the form of a public GitHub repository\footnote{PDC and its extensions, including USuRPER, partial distance correlation periodograms, and the periodogram presented in this work, are all available as part of the SPARTA package \citep{SPARTA2020}, at \url{https://github.com/SPARTA-dev/SPARTA}.}. 

\begin{acknowledgements}We thank the anonymous referee for their wise comments that helped to improve the manuscript.
This research was supported by the Ministry of Innovation, Science \& Technology, Israel (grant 3-18143). The research of SS is supported by a Benoziyo prize postdoctoral
fellowship. 

The analyses done for this paper made use of the code packages: NumPy \citep{numpy}, SciPy \citep{2020SciPy} and SPARTA \citep{SPARTA2020}.
\end{acknowledgements}

\bibliographystyle{aa}
\bibliography{pPDC}

\end{document}